\documentclass[letter,dvipdfmx]{jpsj3}
\usepackage[dvipdfmx]{graphicx}
\usepackage[dvipdfmx]{color}
\renewcommand{\a}{\alpha}

\newcommand{\bea}{\begin{eqnarray}}
\newcommand{\eea}{\end{eqnarray}}
\newcommand{\f}[2]{\frac{#1}{#2}}
\newcommand{\eq}{&=&}
\newcommand{\nn}{\nonumber \\ }
\newcommand{\ve}{\varepsilon}

\newcommand{\area}{\int_{-\infty}^\infty }

\renewcommand{\l}{\lambda}
\newcommand{\p}{\partial}
\newcommand{\pp}[2]{\f{\p #1}{\p #2}}

\newcommand{\siki}[1]{Eq.~(\ref{#1})}
\newcommand{\sikis}[2]{Eqs.~(\ref{#1}) and (\ref{#2})}

\newcommand{\kitai}[1]{\left\langle #1\right\rangle}
\allowdisplaybreaks[1]

\renewcommand{\l}{\lambda}
\renewcommand{\th}{s}

\allowdisplaybreaks[1]
\usepackage{pgf}
\usepackage{tikz}
\usetikzlibrary{arrows,automata}
\usepackage[latin1]{inputenc}
\usepackage{pgfplots}
\usepackage{gnuplot-lua-tikz}
\usepackage{fancyhdr}
\title{
Relationship between optimal portfolios which can maximize and minimize the expected return
}

\author{Takashi Shinzato\thanks{Corresponding author: shinzato@eng.tamagawa.ac.jp}
}
\inst{
Department of Management Science, College of Engineering, 
Tamagawa University, \\ Machida, Tokyo 194-8610, Japan\\
\smallskip
{\rm (Received August 21, 2019; accepted *** 1, 2019)}
} 
\abst{
In recent years, {the evaluation of the minimal investment risk of the quenched disordered system of a portfolio optimization problem and the investment concentration of the optimal portfolio has been actively investigated using the analysis methods of statistical mechanical informatics.} However, the work to date has not sufficiently compared the optimal portfolios of different portfolio optimization problems. Therefore, in this paper, we use the Lagrange undetermined multiplier method and replica analysis to examine the relationship between the optimal portfolios of the expected return maximization problem and the expected return minimization problem with constraints of budget and investment risk. In particular, we derive the mean square error and the correlation coefficient of the optimal portfolios of these maximization and minimization problems as functions of a variable (the degree of risk tolerance) that can characterize the feasible subspace defined by the two constraints.
}
\begin{document}
\maketitle

In recent years, the optimal portfolios and minimal investment risks in the quenched disordered systems of portfolio optimization problems such as the mean-variance model and the absolute-deviation model have been investigated actively using {analysis methods} represented by replica analysis, the belief propagation method, and random matrix theory, which {were} developed in statistical mechanical informatics\cite{
Ciliberti2007,
10.1371/journal.pone.0134968,
KONDOR20071545,
Pafka2002,PAFKA2003487,
doi:10.1080/14697680701422089,
doi:10.1080/1351847X.2011.601661,
110008689817,
1742-5468-2017-12-123402,
1742-5468-2016-12-123404,
doi:10.7566/JPSJ.87.064801,
2018arXiv181006366S,
10.1371/journal.pone.0133846,
PhysRevE.94.062102b,
1742-5468-2017-2-023301,
SHINZATO2018986,
doi:10.7566/JPSJ.86.124804,
Ryosuke-Wakai2014,
doi:10.7566/JPSJ.86.063802,
PhysRevE.94.052307,
1742-5468-2018-2-023401}.
In particular, in operations research, minimal expected investment risk in the annealed disordered system of a portfolio optimization problem and the portfolio {which can minimize the expected investment risk} have been sufficiently discussed, but the investment information required by rational investors is the optimal portfolio of the quenched disordered system of the portfolio optimization problem. For the latter, we need to evaluate the optimal portfolio in the quenched disordered system of the portfolio optimization problem as the true scenario and we also need to reconsider investment problems which have only been analyzed for an annealed disordered system, so that the different methods applied in previous studies can also be used to derive the results expected by rational investors.

However, such previous studies did not sufficiently investigate the properties of a quenched disordered system. In particular, the relationship between the optimal portfolios of quenched disordered systems of portfolio optimization problems has not been sufficiently considered in analyses of quenched and/or annealed disordered systems. Therefore, in this study, we analyze the relationship between the optimal portfolios of a quenched disordered system in a portfolio optimization problem in which multiple constraints are imposed using analysis methods of statistical mechanical informatics.

Note, although the comparison of the optimal portfolios discussed here can be understood intuitively, in order to find the optimal portfolio of the portfolio optimization problem, we use the Boltzmann distribution of the canonical ensemble to compare the ground state in the absolute zero temperature limit from the positive direction and the other ground state in the absolute zero temperature limit from the negative direction (as a matter of convenience), but we do not explicitly discuss both ground states in the absolute zero limits by using the Boltzmann distribution in this paper.
{The aim of this study is} to examine the states at both ends of the canonical ensemble and compare the maximal internal energy state and the minimal internal energy state of the canonical ensemble.

Here, as in previous studies\cite{
doi:10.7566/JPSJ.87.064801,
2018arXiv181006366S,
10.1371/journal.pone.0133846,
PhysRevE.94.062102b,
1742-5468-2017-2-023301,
SHINZATO2018986,
doi:10.7566/JPSJ.86.124804,
Ryosuke-Wakai2014,
doi:10.7566/JPSJ.86.063802,
PhysRevE.94.052307,
1742-5468-2018-2-023401}, we consider the situation of investing in $N$ assets for $p$ periods in an investment market without short-selling regulations and discuss specifically the expected return maximization problem with constraints of budget and investment risk imposed and the properties of the optimal portfolio.
The portfolio of asset $i(=1,2,\cdots,N)$ is $w_i\in{\bf R}$ and 
the portfolio of $N$ assets is denoted
{by} $\vec{w}=(w_1,w_2,\cdots,w_N)^{\rm T}\in{\bf R}^N$, where the notation ${\rm T}$ indicates 
transpose of a vector or matrix.
The return rate of asset $i$ at period $\mu(=1,2,\cdots,p=N\a,\a\sim O(1))$ $\bar{x}_{i\mu}\in{\bf R}$ is independently distributed with 
mean $E[\bar{x}_{i\mu}]=r_i$ and variance $V[\bar{x}_{i\mu}]=v_i$. Then, the expected return of portfolio $\vec{w}$, ${\cal H}(\vec{w})$, is given by
\bea
{\cal H}(\vec{w})\eq\vec{r}^{\rm T}\vec{w},
\eea
where $\vec{r}=(r_1,r_2,\cdots,r_N)^{\rm T}\in{\bf R}^N$ {is used}. Moreover, 
the budget constraint and the investment risk constraint are defined respectively as
\bea
\label{eq2}N\eq\vec{w}^{\rm T}\vec{e},\\
N\ve\eq\f{1}{2}\vec{w}^{\rm T}J\vec{w},
\label{eq3}
\eea
where $\vec{e}=(1,1,\cdots,1)^{\rm T}\in{\bf R}^N$ {is used}; $J=\left\{J_{ij}\right\}\in{\bf R}^{N\times N}$ is the return rate matrix defined by the modified return rate $x_{i\mu}=\bar{x}_{i\mu}-r_i$ having $i,j$ component 
\bea
J_{ij}\eq\f{1}{N}\sum_{\mu=1}^px_{i\mu}x_{j\mu};
\eea
and $\ve$ is the investment risk per asset. 
The latter is related to $\ve_0$, the minimal investment risk per asset of the investment risk minimization problem discussed in previous work \cite{PhysRevE.94.062102b}, by
\bea
\ve\eq \tau\ve_0,
\eea
where $\tau\ge1$
is the degree of risk tolerance. 
As in the same previous work, we use $\ve_0=\f{1}{2g_0}$  and require $p>N$ {in order for $J$ to be} a regular matrix and thus the optimal portfolio to be unique.

\begin{figure}[b]
\centering
\begin{tikzpicture}
      \draw[->,line width=1pt] (-1,0) -- (6.5,0) node[right] {$R$};
      \draw[->,line width=1pt] (0,-1) -- (0,5.5) node[above] {$\ve$};
      \draw[->] (1,4.7)--(1,2.7);
      \draw[->] (5,3.7)--(5,2.7);
      \draw[->] (3,-0.2)--(3,0.5);
\draw[-,line width=2pt](1,2.5)--(5,2.5);

      \draw[domain=0.5:5.5,variable=\x,blue] plot ({\x},{0.5*(1+(\x-3)*(\x-3))}); 

 \draw[domain=-0.2:6,variable=\x,red] plot ({\x},{0.5});
 \draw[domain=-0.2:6,variable=\x,red] plot ({\x},{2.5});
\node at (-0.5,0.5){$\ve_0$};
\node at (3,-0.5){$R=R_1$};
\node at (-0.8,2.5){$\ve=\tau\ve_0$};
\node at (2,5){$R=R_1-\sqrt{V(\tau-1)}$};
\node at (4.6,4){$R=R_1+\sqrt{V(\tau-1)}$};
\end{tikzpicture}
\caption{\label{fig1}$R$ versus $\ve=\tau\ve_0$.
The quadratic function $\ve=\f{1}{2g_0}\left(1+\f{(R-R_1)^2}{V}\right)$ is obtained from $\ve=\tau\ve_0$, $\ve_0=\f{1}{2g_0}$, and 
$\tau=1+\f{(R-R_1)^2}{V}$, which in turn {is obtained from \sikis{eq15-1}{eq16-1}}. The thick line is the expected return per asset in the feasible portfolio subspace satisfying \sikis{eq2}{eq3}.}
\end{figure}
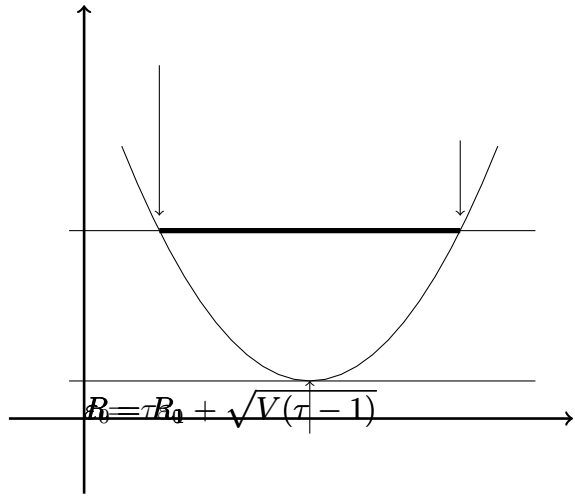

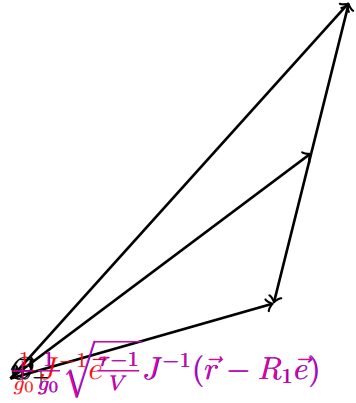
\begin{figure}[bt]
\centering
\begin{tikzpicture}
\draw[->,red,line width=1pt] (0,0) -- (4,3);
\draw[->,magenta,line width=1pt] (4,3) -- (4.5,5);
\draw[->,blue,line width=1pt] (4,3) -- (3.5,1);
\draw[->,line width=1pt] (0,0) -- (4.5,5);
\draw[->,line width=1pt] (0,0) -- (3.5,1);
\node at (2.8,0.3){$\vec{w}_-$};
\node at (2.8,3.7){$\vec{w}_+$};
\node at (-0.2,-0.2){$O$};

\node[red] at (4.8,3){$\f{1}{g_0}J^{-1}\vec{e}$};

\node[blue] at (5.8,1.8){$-\f{1}{g_0}\sqrt{\f{\tau-1}{V}}J^{-1}(\vec{r}-R_1\vec{e})$};

\node[magenta] at (6.3,4.2){$+\f{1}{g_0}\sqrt{\f{\tau-1}{V}}J^{-1}(\vec{r}-R_1\vec{e})$};

\end{tikzpicture}
\caption{\label{fig2}$\vec{w}_+$ and $\vec{w}_-$. Note that vectors
$\vec{w}_+$ and $\vec{w}_-$ are linear combinations of $J^{-1}\vec{e}$ and $J^{-1}(\vec{r}-R_1\vec{e})$. From this figure, it is easy to interpret the correlation coefficients of the $\tau\to1^+$ and $\tau\to\infty$ investment portfolios.  $O$ indicates the origin.} 
\end{figure}

Let us analyze 
the expected return maximization problem with these two constraints using the Lagrange undetermined multiplier method as follows:
\bea
\label{eq6}
L={\cal H}(\vec{w})+k\left(\vec{w}^{\rm T}\vec{e}-N\right)+\th\left(N\ve-\f{1}{2}\vec{w}^{\rm T}J\vec{w}\right),
\eea
where $k$ is the parameter related to budget constraint \siki{eq2} and 
$\th$ is the parameter related to investment risk constraint \siki{eq3}. Then, the extremum obtained from $\pp{L}{w_i}=\pp{L}{k}=\pp{L}{\th}=0$ is given by\bea
\label{eq7}
\vec{w}\eq
\f{1}{\th}J^{-1}\vec{r}+\f{k}{\th}J^{-1}\vec{e},\\
\label{eq8}\th\eq \pm g_0\sqrt{\f{V}{\tau-1}},\\
k\eq\f{\th}{g_0}-R_1,
\eea
where 
\bea
g_0\eq\f{1}{N}\vec{e}^{\rm T}J^{-1}\vec{e},\\
g_1\eq\f{1}{N}\vec{r}^{\rm T}J^{-1}\vec{e},\\
g_2\eq\f{1}{N}\vec{r}^{\rm T}J^{-1}\vec{r},
\eea
and $R_1=\f{g_1}{g_0}$ and
$V=\f{g_2}{g_0}-\left(\f{g_1}{g_0}\right)^2$
are already applied.
Note, the positive term in 
\siki{eq8} $(\th=\th_+)$ is related to 
the maximal expected return, 
and 
the negative term is related to 
the minimal expected return 
as a by-product.
Then, for expected return per asset 
$R=\f{1}{N}{\cal H}(\vec{w})$, the maximal and minimal values $R_+$ and 
$R_-$ are evaluated as follows (see Figure \ref{fig1}):
\bea
\label{eq15-1}
R_+\eq
R_1+\sqrt{V(\tau-1)},\\
\label{eq16-1}R_-\eq
R_1-\sqrt{V(\tau-1)}.
\eea
Therefore,
from \siki{eq7}, 
the two optimal portfolios $\vec{w}_+=\left(w_{1+},w_{2+},\cdots,w_{N+}\right)^{\rm T}$ 
and $\vec{w}_-=\left(w_{1-},w_{2-},\cdots,w_{N-}\right)^{\rm T}\in{\bf R}^N$ 
are given by
\bea
\label{eq16}
\vec{w}_+\eq\f{1}{g_0}\left(J^{-1}\vec{e}+\sqrt{\f{\tau-1}{V}}J^{-1}(\vec{r}-R_1\vec{e})\right),\\
\label{eq17}\vec{w}_-\eq\f{1}{g_0}\left(J^{-1}\vec{e}-\sqrt{\f{\tau-1}{V}}J^{-1}(\vec{r}-R_1\vec{e})\right).
\eea
As shown in Figure \ref{fig2},
these are linear combinations of the constant vectors 
$J^{-1}\vec{e}$, {whose coefficient} is independent of $\tau$, and $J^{-1}(\vec{r}-R_1\vec{e})$, {whose coefficient} depends on $\tau$. 
Note that {the second} vector $J^{-1}(\vec{r}-R_1\vec{e})$ is always orthogonal to $\vec{e}$.

An additional investment index, besides the expected return,
is the investment concentration $q_w=\f{1}{N}\vec{w}^{\rm T}\vec{w}$.
The investment concentration of the maximal expected return $q_{w+}=\f{1}{N}{\vec{w}_+}^{\rm T}\vec{w}_+$ and that 
of the minimal expected return $q_{w-}=\f{1}{N}{\vec{w}_-}^{\rm T}\vec{w}_-$ are respectively calculated as follows:
\bea
q_{w+}=\f{f_0(\tau-1)}{g_0^2V}
\left(V_f+\left(\f{f_1}{f_0}-\f{g_1}{g_0}+\sqrt{\f{V}{\tau-1}}\right)^2\right),
\\
q_{w-}=\f{f_0(\tau-1)}{g_0^2V}
\left(V_f+\left(\f{f_1}{f_0}-\f{g_1}{g_0}-\sqrt{\f{V}{\tau-1}}\right)^2\right),
\eea
where $V_f=\f{f_2}{f_0}-\left(\f{f_1}{f_0}\right)^2$ 
and 
\bea
f_0\eq\f{1}{N}\vec{e}^{\rm T}J^{-2}\vec{e},\\
f_1\eq\f{1}{N}\vec{r}^{\rm T}J^{-2}\vec{e},\\
f_2\eq\f{1}{N}\vec{r}^{\rm T}J^{-2}\vec{r}.
\eea
The overlap between $\vec{w}_+$ 
and $\vec{w}_-$, $c=\f{1}{N}{\vec{w}_+}^{\rm T}\vec{w}_-$, 
given by $c=-\f{f_0(\tau-1)}{g_0^2V}
\left(V_f+\left(\f{f_1}{f_0}-\f{g_1}{g_0}\right)^2-\f{V}{\tau-1}\right)$, 
is introduced for simplicity, and then 
the mean square error of $\vec{w}_+$ and $\vec{w}_-$, 
$\Delta=\f{1}{N}\sum_{i=1}^N(w_{i+}-w_{i-})^2
=q_{w+}+q_{w-}-2c$, 
\bea
\label{eq-delta}
\Delta
\eq\f{4f_0(\tau-1)}{g_0^2V}\left(V_f+\left(\f{f_1}{f_0}-\f{g_1}{g_0}\right)^2\right),
\eea
is evaluated. From 
\siki{eq-delta}, 
the mean square error of 
$\vec{w}_+$ and $\vec{w}_-$, $\Delta$, is 
proportional to $\tau-1$,
and so, as shown in Figure \ref{fig2}, as $\tau\to1^+$,
$\Delta$ approaches 0.

Moreover, so as to compare the 
geometric relationship between the optimal portfolios, 
the correlation coefficient between $\vec{w}_+$ and $\vec{w}_-$, 
$\rho=\f{\vec{w}_+^{\rm T}\vec{w}_-}{
\sqrt{|\vec{w}_+|^2
|\vec{w}_-|^2
}
}$, is calculated as follows (note that the angle between the two vectors is $\arccos\rho$):
\bea
\rho
\eq-\f{A}{
\sqrt{
A^2
+4\f{V}{\tau-1}
\left(\f{f_2}{f_0}-\left(\f{f_1}{f_0}\right)^2\right)
}},\label{eq26}
\eea
where here 
$A=V_f+\left(\f{f_1}{f_0}-\f{g_1}{g_0}\right)^2-\f{V}{\tau-1}$ 
is applied. 
From this result, as $\tau$ becomes large, $\rho$ approaches $-1$. 
From \sikis{eq16}{eq17}, 
for sufficiently large $\tau$, 
the second term in both equations, $J^{-1}(\vec{r}-R_1\vec{e})$, 
is larger than first term, $J^{-1}\vec{e}$; 
that is, the second term dominates.
In a similar way,
when $\tau\to1^+$,
$\rho$ approaches 1.
From \sikis{eq16}{eq17}, we can interpret this as meaning that $\vec{w}_+\simeq\vec{w}_-$. 
Here the following additional point should be noted.
The discussion involving from \siki{eq6} to \siki{eq26} does not consider 
the statistical properties of the modified return rate $x_{i\mu}$, 
so it holds for an arbitrary statistical distribution of the return rate.

The statistical properties of the return are included in the values of 
$g_0,g_1,g_2,f_0,f_1,$ and $f_2$ 
calculated using the inverse of the return rate matrix, $J^{-1}$,
in the above discussion, but we have not yet evaluated them.
Although we postulate that in this case
we can analyze $g_0,g_1,g_2,f_0,f_1$, and $f_2$, 
in general 
the computational complexity of solving the inverse matrix $J^{-1}$ is $O(N^3)$.
When the number of assets, $N$, is large, these six moments are difficult to analyze.
For this reason, we do not directly assess $J^{-1}$ below,
but instead consider analyzing 
$g_0,g_1,g_2,f_0,f_1,$ and $f_2$ 
by using replica analysis.

Similar to in previous work, we define the partition function as follows:
\bea
\label{eq26-1}Z\eq
\area 
\f{d\vec{w}}{(2\pi)^{\f{N}{2}}}
e^{-\f{1}{2}\vec{w}^{\rm T}(J-\l I)\vec{w}+k\vec{w}^{\rm T}\vec{e}+\theta\vec{r}^{\rm T}\vec{w}}
,
\eea
where $I\in{\bf R}^{N\times N}$ is the identity matrix. 
The integral in \siki{eq26-1} can be evaluated with respect to $\vec{w}$ and gives 
$\phi(\l)=\f{1}{N}\log Z$, 
where\bea
\phi(\l)
\eq\f{1}{2N}
\left(k\vec{e}+\theta\vec{r}\right)^{\rm T}
(J-\l I)^{-1}\left(k\vec{e}+\theta\vec{r}\right)
\nn
&&-\f{1}{2N}\log\det\left|J-\l I\right|.
\eea
If we then partially differentiate $\phi(\l)$ with respect to 
$\l$,
\bea
\phi'(\l)\eq\f{1}{2N}{\rm Tr}
\left\{
(J-\l I)^{-2}\left(k\vec{e}+\theta\vec{r}\right)\left(k\vec{e}+\theta\vec{r}\right)^{\rm T}
\right\}\nn
&&+\f{1}{2N}{\rm Tr}(J-\l I)^{-1}
\eea
is obtained. From this, when $\l=0$, $\phi(0)$ and $\phi'(0)$ are given as follows:
\bea
\phi(0)\eq\f{k^2}{2}\f{\vec{e}^{\rm T}J^{-1}\vec{e}}{N}+k\theta\f{\vec{r}^{\rm T}J^{-1}\vec{e}}{N}+\f{\theta^2}{2}
\f{\vec{r}^{\rm T}J^{-1}\vec{r}}{N}\nn
&&-\f{1}{2N}\log \det\left|J\right|\nn
\eq\f{k^2}{2}g_0+k\theta g_1+\f{\theta^2}{2}g_2-\f{1}{2N}\log\det|J|,\quad\\
\phi'(0)\eq
\f{k^2}{2}\f{\vec{e}^{\rm T}J^{-2}\vec{e}}{N}+k\theta\f{\vec{r}^{\rm T}J^{-2}\vec{e}}{N}+\f{\theta^2}{2}
\f{\vec{r}^{\rm T}J^{-2}\vec{r}}{N}\nn
&&+\f{1}{2N}{\rm Tr}J^{-1}\nn
\eq\f{k^2}{2}f_0+k\theta f_1+\f{\theta^2}{2}f_2+\f{1}{2N}{\rm Tr}J^{-1}.
\eea
If we then  partially differentiate $\phi(0)$ and $\phi'(0)$
with respect to $k$ and $\theta$, we can derive $g_0,g_1,g_2,f_0,f_1,$ and $f_2$. We apply this strategy for evaluating the six moments in the following discussion.

Using replica analysis, let us first evaluate $\phi(\l)$. For $n\in{\bf Z}$,
based on the self-averaging property of the configuration average of $Z^n$, considering $E[Z^n]$ in the limit of a large number of assets $N$, we evaluate
\bea
&&
\lim_{N\to\infty}\f{1}{N}\log E[Z^n]\nn
\eq\mathop{\rm Extr}_{Q_s,\tilde{Q}_s}\left\{
-\f{\a}{2}\log\det\left|I+Q_s\right|+\f{1}{2}{\rm Tr}Q_s\tilde{Q}_s\right.\nn
&&
-\f{1}{2}
\kitai{\log\det\left|v\tilde{Q}_s-\l I\right|}\nn
&&
\left.
+\f{1}{2}
\kitai{(k+\theta r)^2\vec{e}^{\rm T}(v\tilde{Q}_s-\l I)^{-1}\vec{e}}
\label{eq31}
\right\},
\eea
where $\a=p/N\sim O(1)$, $I\in{\bf R}^{n\times n}$ is the identity matrix, and 
the order parameter matrix $Q_s=\left\{q_{sab}\right\}\in{\bf R}^{n\times n}$ 
and conjugate parameter matrix $\tilde{Q}_s=\left\{\tilde{q}_{sab}\right\}\in{\bf R}^{n\times n}$ 
are already used,
as well as the following notation:
\bea
\kitai{f(r,v)}
\eq\lim_{N\to\infty}\f{1}{N}
\sum_{i=1}^Nf(r_i,v_i).
\eea
Furthermore, the notation $\mathop{\rm Extr}_xg(x)$ means 
the extremum of function $g(x)$ 
with respect to the variable $x$.

Now, when $\l=0$, from the extremum of \siki{eq31},
\bea
\label{eq33}
\tilde{Q}_s\eq\a(I+Q_s)^{-1},\\
Q_s\eq
\tilde{Q}_s^{-1}+
\kitai{\f{(k+\theta r)^2}{v}}
\tilde{Q}_s^{-1}D
\tilde{Q}_s^{-1}\label{eq34}
\eea
is obtained, where $D\in{\bf R}^{n\times n}$ is the square matrix {whose components are all 1}.Moreover, since $Q_s,\tilde{Q}_s$ in \sikis{eq33}{eq34} can be described as linear combinations of 
$I$ and $D$, that is, we can set 
$Q_s=\chi_sI+q_sD,
\tilde{Q}_s=\tilde{\chi}_sI-\tilde{q}_sD$,
the results of replica analysis obtained as $n\to0$ are  \bea
\label{eq35}
\chi_s\eq\f{1}{\a-1},\\
q_s\eq\f{\a}{(\a-1)^3}\kitai{\f{(k+\theta r)^2}{v}},\\
\tilde{\chi}_s\eq\a-1,\\
\label{eq38}
\tilde{q}_s\eq\f{1}{\a-1}
\kitai{\f{(k+\theta r)^2}{v}}.
\eea
Note that although we did not assume that
$Q_s$ and $\tilde{Q}_s$ in \siki{eq31} are given by the replica symmetry solution, since the solutions \sikis{eq33}{eq34} satisfy  
$Q_s=\chi_sI+q_sD$ and 
$\tilde{Q}_s=\tilde{\chi}_sI-\tilde{q}_sD$,
it turns out that the replica symmetry solution is actually the exact solution.

From the above, $\phi(\l)=\lim_{n\to0}\pp{}{n} 
\lim_{N\to\infty}\f{1}{N}\log E[Z^n]
$ is
\bea
\phi(\l)
\eq
-\f{\a}{2}\log(1+\chi_s)-\f{\a q_s}{2(1+\chi_s)}
\nn
&&+\f{1}{2}(\chi_s+q_s)(\tilde{\chi}_s-\tilde{q}_s)
+\f{q_s\tilde{q}_s}{2}\nn
&&-\f{1}{2}
\kitai{\log(v\tilde{\chi_s}-\l)}
+\f{1}{2}\kitai{\f{v\tilde{q}_s}{v\tilde{\chi}_s-\l}}\nn
&&+\f{1}{2}\kitai{\f{(k+\theta r)^2}{v\tilde{\chi}_s-\l}}.
\eea
Therefore, substituting 
the results from Eqs. (\ref{eq35}) to (\ref{eq38}) into $\phi(0)$ and $\phi'(0)$ gives 
\bea
\phi(0)\eq-\f{\a}{2}\log(1+\chi_s)-\f{\a q_s}{2(1+\chi_s)}\nn
&&+\f{1}{2}(\chi_s+q_s)(\tilde{\chi}_s-\tilde{q}_s)+\f{q_s\tilde{q}_s}{2}
-\f{1}{2}\kitai{\log v}\nn
&&-\f{1}{2}\log\tilde{\chi}_s+\f{\tilde{q}_s}{2\tilde{\chi}_s}+\f{1}{2\tilde{\chi}_s}\kitai{v^{-1}(k+\theta r)^2}\nn
\eq -\f{\a}{2}\log\a+\f{\a+1}{2}\log(\a-1)
+\f{1}{2}\nn
&&-\f{1}{2}\kitai{\log v}
+\f{1}{2(\a-1)}\kitai{v^{-1}(k+\theta r)^2},\\
\phi'(0)
\eq\f{1}{2\tilde{\chi}_s}\kitai{v^{-1}}
+\f{\tilde{q}_s}{2\tilde{\chi}_s^2}\kitai{v^{-1}}+
\f{1}{2\tilde{\chi}_s^2}
\kitai{\f{(k+\theta r)^2}{v^2}}\nn
\eq\f{1}{2(\a-1)}\kitai{v^{-1}}
+\f{\kitai{v^{-1}}}{2(\a-1)^3}
\kitai{v^{-1}(k+\theta r)^2}\nn
&&
+\f{1}{2(\a-1)^2}
\kitai{v^{-2}(k+\theta r)^2}.
\eea
Finally, from 
$g_0=\pp{^2\phi(0)}{k^2},
g_1=\pp{^2\phi(0)}{k\p \theta},
g_2=\pp{^2\phi(0)}{\theta^2},
f_0=\pp{^2\phi'(0)}{k^2},
f_1=\pp{^2\phi'(0)}{k\p \theta},$ and 
$f_2=\pp{^2\phi'(0)}{\theta^2}
$,
\bea
g_0\eq\f{\kitai{v^{-1}}}{\a-1},\\
g_1\eq\f{\kitai{v^{-1}r}}{\a-1},\\
g_2\eq\f{\kitai{v^{-1}r^2}}{\a-1},\\
f_0\eq\f{\kitai{v^{-1}}^2}{(\a-1)^3}+\f{\kitai{v^{-2}}}{(\a-1)^2},\\
f_1\eq\f{\kitai{v^{-1}}\kitai{v^{-1}r}}{(\a-1)^3}+\f{\kitai{v^{-2}r}}{(\a-1)^2},\\
f_2\eq\f{\kitai{v^{-1}}\kitai{v^{-1}r^2}}{(\a-1)^3}+\f{\kitai{v^{-2}r^2}}{(\a-1)^2},
\eea
are obtained. From this result, 
\bea
R_1\eq\f{\kitai{v^{-1}r}}{\kitai{v^{-1}}},\\
V\eq
\f{\kitai{v^{-1}r^2}}{\kitai{v^{-1}}}
-\left(\f{\kitai{v^{-1}r}}{\kitai{v^{-1}}}\right)^2,
\eea
are also obtained. 
Thus, we succeeded in evaluating analytically maximal expected return per asset $R_+$ and minimal expected return per asset $R_-$ in 
\sikis{eq15-1}{eq16-1}, 
and optimal portfolios $\vec{w}_+$ and $\vec{w}_-$ in 
\sikis{eq16}{eq17}.

In this paper, we have discussed the expected return maximization problem with constraints of budget and investment risk, as well as the expected return minimization problem as a by-product. In particular, we reformulated the expected return maximization problem with two constraints by using the Lagrange undetermined multiplier method. As a result, maximal expected return per asset $R_+$ with corresponding optimal portfolio $\vec{w}_+$ and minimal expected return per asset $R_-$ with corresponding optimal portfolio $\vec{w}_-$ were successfully derived. Moreover, $\Delta$, which is the mean square error of $\vec{w}_+$ and 
$\vec{w}_-$, and the correlation coefficient $\rho$ were analytically derived.
We were also able to briefly interpret the correlation coefficient for both $\tau\to\infty$ and $\tau\to1^+$ by 
using Figure \ref{fig2}.
In addition, 
using replica analysis, we could analytically evaluate the six moments from the Lagrange undetermined multiplier method. 
In particular,
with respect to order parameters $q_{sab},\tilde{q}_{sab}$,
although we do not use the ansatz of replica symmetry {solution},
from saddle point equations \sikis{eq34}{eq35},
it is shown that, 
automatically, the assumption of the replica symmetry solution is validated.

In this study, we mainly deal with the portfolio optimization problem when the return rates of assets are independently distributed, but the discussion involving from Eq. (\ref{eq6}) to (\ref{eq26}) holds for any distribution of the return rate. Therefore, in order to meet the expectations of rational investors, it is necessary to evaluate the six moments even when the return rates of assets are correlated with each other and to extend the theoretical analysis method which we here propose. There is also a need to compare the optimal portfolios for portfolio optimization problems other than the expected {return} maximization problem.
Especially, 
with respect to the optimal portfolios of the investment concentration maximization/minimization problems with constraints of budget and investment risk
\cite{SHINZATO2018986,doi:10.7566/JPSJ.86.124804}, we need to analyze the relationship between both ends of the canonical ensemble.

The author is grateful for fruitful discussions with I.~Suzuki and D.~Tada. 
This work was partially supported by
Grants-in-Aid Nos.~15K20999, 17K01260, and 17K01249; Research Project of the Institute of Economic Research Foundation at Kyoto University; and Research Project
No.~4 of the Kampo Foundation.

\bibliographystyle{jpsj}
\bibliography{sample20190105}

\end{document}